\documentclass[12pt,a4paper]{iopart}
\usepackage{iopams}
\usepackage{graphicx} 
\usepackage[caption=false]{subfig}
\usepackage{pdfpages} 
\begin{document}


\title{Grover Search with Lackadaisical Quantum Walks}

\author{Thomas G Wong}
\address{Faculty of Computing, University of Latvia, Rai\c{n}a bulv.~19, R\=\i ga, LV-1586, Latvia}
\ead{\mailto{twong@lu.lv}}

\begin{abstract}
	The lazy random walk, where the walker has some probability of staying put, is a useful tool in classical algorithms. We propose a quantum analogue, the \emph{lackadaisical} quantum walk, where each vertex is given $l$ self-loops, and we investigate its effects on Grover's algorithm when formulated as search for a marked vertex on the complete graph of $N$ vertices. For the discrete-time quantum walk using the phase flip coin, adding a self-loop to each vertex boosts the success probability from $1/2$~to~$1$. Additional self-loops, however, decrease the success probability. Using instead the Shenvi, Kempe, and Whaley (2003) coin, adding self-loops simply slows down the search. These coins also differ in that the first is faster than classical when $l$ scales less than $N$, while the second requires that $l$ scale less than $N^2$. Finally, continuous-time quantum walks differ from both of these discrete-time examples---the self-loops make no difference at all. These behaviors generalize to multiple marked vertices.
\end{abstract}



\section{Introduction}

Random walks, or Markov chains, are the basis for a variety of classical algorithms \cite{Norris1998}. One typically wants the random walker to move readily, but there are some scenarios when it is desirable for the walker to be \emph{lazy}, meaning it has some probability of staying put. For example, if a normal (non-lazy) random walker starts in one of the two vertex sets of a bipartite graph, then at each time step, the walker will only be in one vertex set. By making the walk lazy, however, the walker can have probability of being in both vertex sets, improving its coverage of the graph. Such lazy random walks equate to adding self-loops to the vertices of the graph with appropriate weight, and they have been utilized in a variety of classical algorithms, including PageRank \cite{ACL2006}, graph covering \cite{AKL2008}, and image processing \cite{SDWL2014}.

Given the success of lazy random walks in the classical regime, we propose in this paper a quantum analogue called \emph{lackadaisical} quantum walks, defined to be a standard quantum walk \cite{ADZ1993,Ambainis2003,Kempe2003} on a graph with the addition of $l$ self-loops to each vertex of the graph. So the greater $l$ is, the more the walker prefers to stay put. Note this differs from the ``lazy'' quantum walk proposed by \cite{Childs2010}, hence our choice of a different name. It also generalizes three-state lazy quantum walks on the line \cite{IKS2005,FB2014,SBJ2014,Dan2015}, which only have one self-loop at each vertex. Since there exist both discrete- and continuous-time varieties of quantum walks \cite{Kempe2003}, we will consider both in our investigation of lackadaisical quantum walks.

In particular, we explore the addition of self-loops to one of the best-known problems in computing: unstructured search, whose famous quantum solution is Grover's algorithm \cite{Grover1996}. To review, given a ``database'' with entries $1, 2, \dots, N$, and an oracle $f(x)$ that outputs $1$ for a particular ``marked'' entry $w$ and $0$ otherwise, a classical computer expects to query the oracle $N/2 = \Theta(N)$ times before finding $w$, since it could be the first guess or the last. In the quantum setting, $| 1 \rangle, | 2 \rangle, \dots, | N \rangle$ are computational basis states, and the oracle $R_w = (-1)^{f(x)}$ acts by flipping the phase of a marked basis state $| w \rangle$ while leaving the rest unchanged, \textit{i.e.}, $R_w | w \rangle = -| w \rangle$ and $R_w | x \rangle = | x \rangle$, $\forall x \ne w$. This reflection through $| w \rangle$ can be written as $R_w = I - 2 | w \rangle \langle w |$. Grover's algorithm \cite{Grover1996} finds $| w \rangle$ in only $\Theta(\sqrt{N})$ queries, which is a quadratic speedup over the classical $\Theta(N)$, and it does so by initializing the system in an equal superposition
\[ | s \rangle = \frac{1}{\sqrt{N}} \sum_{i=1}^N | i \rangle \]
over the basis states and repeatedly applying
\begin{equation}
	\label{eq:Grover}
	R_{s^\perp} R_w,
\end{equation}
where $R_{s^\perp} = 2 | s \rangle \langle s | - I$ is a reflection through $| s^\perp \rangle$. Applying these two reflections $\pi\sqrt{N}/4$ times, the state is rotated from $| s \rangle$ to $| w \rangle$ with probability near $1$.

\begin{figure}
\begin{center}
	\subfloat[]{
		\includegraphics{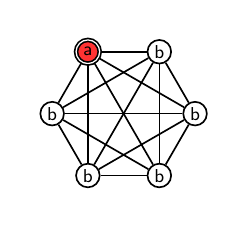}
		\label{fig:complete}
	} \quad
	\subfloat[] {
		\includegraphics{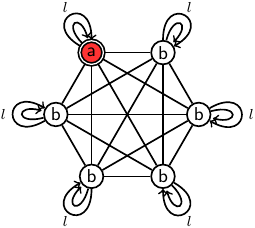}
		\label{fig:complete_loops}
	}
	\caption{(a) The complete graph with $N = 6$ vertices. A vertex is marked, as indicated by a double circle. Identically evolving vertices are identically colored and labeled, and the labels indicate the subspace basis vectors that the vertices belong to. (b) With $l$ self-loops at each vertex.}
\end{center}
\end{figure}

As a graph problem, we formulate this unstructured search problem as search on the complete graph of $N$ vertices for a particular marked vertex, an example of which is shown in \fref{fig:complete}. Since each vertex is connected to every other, there is no structure demanding an order to which we visit the vertices. Thus a classical random walk that jumps from vertex to vertex, checking at each step if it has found $w$, expects to make $\Theta(N)$ such steps and checks to find the marked vertex. A quantum walk, on the other hand, searches in Grover's $\Theta(\sqrt{N})$, and we explicitly show this in the next two sections in both discrete and continuous time \cite{CG2004}. Then we add $l$ self-loops to each vertex, as shown in \fref{fig:complete_loops}, and show how the lackadaisical walk affects the search algorithms. For the discrete-time quantum walk, we get an improvement in the success probability of the algorithm with a particular ``coin'' with $l = 1$. Additional self-loops, or search with a different coin, hurts the algorithm. The continuous-time algorithm, on the other hand, is not affected by the self-loops at all. Finally, we generalize the results to multiple marked vertices.


\section{Grover's Algorithm as a Discrete-Time Quantum Walk}

We begin by analyzing quantum walk search on the complete graph with no self-loops, in discrete-time in this section and continuous-time in the next. For both discrete- and continuous-time quantum walks, the quantum walker jumps from vertex to vertex, and the vertices of the graph label computational basis states of an $N$-dimensional ``vertex'' Hilbert space $\mathbb{C}^N$. For discrete-time quantum walks, however, this is insufficient to define a local unitary operator \cite{Meyer1996a,Meyer1996b}, so we necessarily include an additional $d$-dimensional ``coin'' Hilbert space $\mathbb{C}^d$ supported by the directions/edges that the particle can jump along from each vertex. For the complete graph, each vertex is connected to the other $N-1$ vertices, so $d = N-1$. Let $| s_v \rangle$ and $| s_c \rangle$ be equal superpositions over each space:
\[ | s_v \rangle = \frac{1}{\sqrt{N}} \sum_{i=1}^N | i \rangle, \quad | s_c \rangle = \frac{1}{\sqrt{d}} \sum_{i=1}^{d} | i \rangle. \]
Then the system $| \psi \rangle$ begins in the equal superposition over the entire $\mathbb{C}^N \otimes \mathbb{C}^{d}$ Hilbert space:
\[ | \psi_0 \rangle = | s_v \rangle \otimes | s_c \rangle. \]
The quantum walk is defined by repeated applications of
\[ U_0 = S \cdot \left( I_N \otimes C_0 \right), \]
where $C_0$ is the ``Grover diffusion'' coin \cite{SKW2003}
\[ C_0 = 2 | s_c \rangle \langle s_c | - I_d, \]
and $S$ is the ``flip-flop'' shift \cite{AKR2005} that causes a particle to hop and then turn around, \textit{e.g.}, a particle at vertex $1$ that points towards vertex $2$ will, after an application of $S$, be at vertex $2$ and point towards vertex $1$: $S (|1\rangle \otimes |1 \to 2\rangle) = |2\rangle \otimes |2 \to 1\rangle$. Note that $| \psi_0 \rangle$ is the equilibrium distribution of this walk, so $U_0 | \psi_0 \rangle = | \psi_0 \rangle$.

To turn this into a search algorithm, we apply a different coin $C_1$ to the marked vertex and still use $C_0$ on the unmarked vertices \cite{SKW2003}, so the search operator is
\begin{equation}
	\label{eq:U}
	U = S \cdot \left[ \left(I_N - | w \rangle \langle w | \right) \otimes C_0 + | w \rangle \langle w | \otimes C_1 \right].
\end{equation}
Two choices for $C_1$ are common \cite{AKR2005}. The first is $C_1^{\rm flip} = -C_0$, which causes $U$ to become
\begin{eqnarray}
	U
	&= S \cdot \left[ \left(I_N - 2 | w \rangle \langle w | \right) \otimes C_0 \right] \nonumber \\
	&= \underbrace{S \cdot \left( I_N \otimes C_0 \right)}_{U_0} \cdot \underbrace{ \left(I_N - 2 | w \rangle \langle w | \right)}_{R_w} \otimes I_d \nonumber \\
	&= U_0 \cdot (R_w \otimes I_d) \label{eq:Uflip}.
\end{eqnarray}
Note $R_w$ is the phase flip in Grover's algorithm, so this applies a phase flip to the marked vertex followed by a step of the quantum walk. The second common choice for $C_1$ is $C_1^{\rm SKW} = -I_d$, which was first introduced by Shenvi, Kempe, and Whaley \cite{SKW2003} to solve search on the hypercube, and later used by Ambainis, Kempe, and Rivosh \cite{AKR2005} to solve search on arbitrary dimensional periodic square lattices.

\begin{figure}
\begin{center}
	\includegraphics{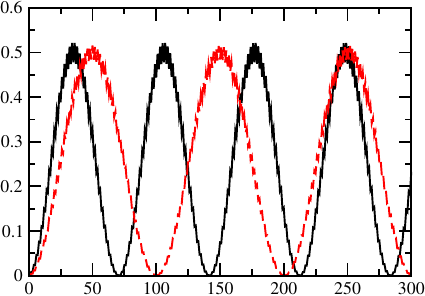}
	\caption{\label{fig:discrete_l0} Success probability as a function of the number of discrete-time applications of $U$ (with either the $C_1^{\rm flip}$ or $C_1^{\rm SKW}$ coins) for search on the complete graph with $N = 1024$ (solid black) and $2048$ (dashed red) vertices.}
\end{center}
\end{figure}

With either of these coins, the system evolves such that there are only two types of vertices, as shown in \fref{fig:complete}. In particular, the marked red $a$ vertex evolves differently from the identically-evolving unmarked white $b$ vertices. Since the $a$ vertex can only point towards $b$ vertices, and the $b$ vertices can either point towards the $a$ vertex or other $b$ vertices, the system evolves in a 3D subspace, and we take equal superpositions of these vertices/directions as the basis vectors:
\begin{eqnarray*}
	| ab \rangle = | a \rangle \otimes \frac{1}{\sqrt{N-1}} \sum_b | a \to b \rangle \\
	| ba \rangle = \frac{1}{\sqrt{N-1}} \sum_b | b \rangle \otimes | b \to a \rangle \\
	| bb \rangle = \frac{1}{\sqrt{N-1}} \sum_b | b \rangle \otimes \frac{1}{\sqrt{N-2}} \sum_{b' \sim b} | b \to b' \rangle.
\end{eqnarray*}
In this $\{ | ab \rangle, | ba \rangle, | bb \rangle \}$ basis, the initial state is
\[ | \psi_0 \rangle = \frac{1}{\sqrt{N}} \left( | ab \rangle + | ba \rangle + \sqrt{N-2} | bb \rangle \right). \]
In general, the search operator \eref{eq:U} is different for $C_1^{\rm flip}$ and $C_1^{\rm SKW}$. But for the complete graph, they are identical; with either coin, the search operator is
\begin{equation}
	\label{eq:U_noloops}
	U =  \left( \!\!\! \begin{array}{ccc}
		0 & -\cos \theta & \sin \theta \\
		-1 & 0 & 0 \\
		0 & \sin \theta & \cos \theta \\
	\end{array} \! \right),
\end{equation}
where
\[ \cos \theta = \frac{N-3}{N-1}, \quad {\rm and} \quad \sin \theta = \frac{2\sqrt{N-2}}{N-1}. \]
Repeatedly applying this operator to the initial state, the success probability evolves as shown in \fref{fig:discrete_l0} for $N = 1024$ and $2048$ vertices. We see that the success probability reaches $1/2$, and the runtime (\textit{i.e.}, number of applications of $U$ to reach the maximum success probability) scales less than linear (\textit{i.e.}, classical) since doubling $N$ results in a runtime that is less than double. In particular, we expect it to scale as $\Theta(\sqrt{N})$ to be a formulation of Grover's algorithm.

To prove this behavior and find the precise runtime, we first find the eigenvectors and eigenvalues of $U$. They are
\[ \left| \psi_{+} \right\rangle = \sqrt{\frac{1+\cos\theta}{3+\cos\theta}} \left( \!\! \begin{array}{c}
	\frac{1}{2\sqrt{1+\cos\theta}} \left[ \sqrt{1-\cos\theta} - i \sqrt{3+\cos\theta} \right] \\
	\frac{1}{2\sqrt{1+\cos\theta}} \left[ \sqrt{1-\cos\theta} + i \sqrt{3+\cos\theta} \right] \\
	1 \\
\end{array} \! \right), \enspace E_+ = \rme^{\rmi\phi} \]
\[ \left| \psi_{-} \right\rangle = \sqrt{\frac{1+\cos\theta}{3+\cos\theta}} \left( \!\! \begin{array}{c}
	\frac{1}{2\sqrt{1+\cos\theta}} \left[ \sqrt{1-\cos\theta} + i \sqrt{3+\cos\theta} \right] \\
	\frac{1}{2\sqrt{1+\cos\theta}} \left[ \sqrt{1-\cos\theta} - i \sqrt{3+\cos\theta} \right] \\
	1 \\
\end{array} \! \right), \enspace E_- = \rme^{-\rmi\phi} \]
\[ \left| \psi_{-1} \right\rangle = \sqrt{\frac{1-\cos\theta}{3+\cos\theta}} \left( \!\! \begin{array}{c}
	-\sqrt{\frac{1+\cos\theta}{1-\cos\theta}} \\
	-\sqrt{\frac{1+\cos\theta}{1-\cos\theta}} \\
	1 \\
\end{array} \right), \enspace E_{-1} = -1 \]
where $\phi$ is defined such that
\[ \cos\phi = \frac{1+\cos\theta}{2}, \quad {\rm and} \quad \sin\phi = \frac{\sqrt{(1-\cos\theta)(3+\cos\theta)}}{2}. \]

Now we express the initial state in terms of the eigenvectors of $U$. Consider
\[ \frac{1}{\sqrt{2}} \left( \left| \psi_+ \right\rangle + \left| \psi_- \right\rangle \right) = \frac{1}{\sqrt{2}} \sqrt{\frac{1+\cos\theta}{3+\cos\theta}} \left( \!\! \begin{array}{c}
	\sqrt{\frac{1-\cos\theta}{1+\cos\theta}} \\
	\sqrt{\frac{1-\cos\theta}{1+\cos\theta}} \\
	2 \\
\end{array} \right). \]
For large $N$, $\sin\theta \approx 2/\sqrt{N}$ implies that $\theta \approx 2/\sqrt{N}$, so the first two components of this are
\[ \sqrt{\frac{1-\cos\theta}{1+\cos\theta}} \approx \sqrt{\frac{\theta^2/2}{2}} = \sqrt{\frac{\theta^2}{4}} = \frac{\theta}{2} \approx \frac{1}{\sqrt{N}}, \]
which means the last term dominates for large $N$. That is,
\[ | bb \rangle \approx \frac{1}{\sqrt{2}} \left( \left| \psi_+ \right\rangle + \left| \psi_- \right\rangle \right). \]
Since $| \psi_0 \rangle \approx | bb \rangle$, the system after $t$ applications of $U$ is
\[ U^t | \psi_0 \rangle \approx \frac{1}{\sqrt{2}} \left( U^t \left| \psi_+ \right\rangle + U^t \left| \psi_- \right\rangle \right) = \frac{1}{\sqrt{2}} \left( \rme^{\rmi \phi t} \left| \psi_+ \right\rangle + \rme^{-\rmi \phi t} \left| \psi_- \right\rangle \right). \]
When $\phi t = \pi/2$, \textit{i.e.},
\[ t = \frac{\pi}{2\phi} = \frac{\pi}{2 \sin^{-1} \left( \frac{\sqrt{(1-\cos\theta)(3+\cos\theta)}}{2} \right)} \approx \frac{\pi}{2 \sin^{-1} (\theta/\sqrt{2})} \approx \frac{\pi}{2\sqrt{2}} \sqrt{N}, \]
the state of the system is approximately
\[ \frac{1}{\sqrt{2}} \left( i \left| \psi_+ \right\rangle - i \left| \psi_- \right\rangle \right) = \frac{1}{\sqrt{2}} \sqrt{\frac{1+\cos\theta}{3+\cos\theta}} \left( \!\!\! \begin{array}{c}
	\sqrt{\frac{3+\cos\theta}{1+\cos\theta}} \\
	-\sqrt{\frac{3+\cos\theta}{1+\cos\theta}} \\
	0 \\
\end{array} \! \right)
= \frac{1}{\sqrt{2}} \left( \!\! \begin{array}{c}
	1 \\
	-1 \\
	0 \\
\end{array} \!\!\! \right). \]
So the system roughly evolves from $| bb \rangle$ to being half in $| ab \rangle$ and half in $| ba \rangle$, which from the $| ab \rangle$ component gives a success probability of $1/2$. This agrees with \fref{fig:discrete_l0}; the success probability reaches $1/2$ after $\pi\sqrt{1024}/2\sqrt{2} \approx 36$ and $\pi\sqrt{2048}/2\sqrt{2} \approx 50$ applications of $U$. Repeating the algorithm an expected constant number of times to boost the success probability near $1$, the algorithm still finds the marked vertex in $\Theta(\sqrt{N})$ applications of $U$, which is the same scaling as Grover's algorithm.

This explicit proof that the success probability reaches $1/2$ in $\pi\sqrt{N}/2\sqrt{2}$ applications of $U$ seems original, even though it is well-known that the success probability reaches $1$ in continuous-time \cite{CG2004} (which we review next), or in discrete-time with one self-loop per vertex \cite{AKR2005} (which we review later). Note that the first discrete-time quantum walk search algorithm \cite{SKW2003} was search on the hypercube, not the complete graph.


\section{Grover's Algorithm as a Continuous-Time Quantum Walk}

In continuous-time, quantum walk search on the complete graph with no self-loops ($l = 0$) was previously investigated by \cite{CG2004}, and we review the results here. Continuous-time quantum walks do not require the coin space, so they walk in the $N$-dimensional Hilbert space $\mathbb{C}^N$ supported by the vertices of the graph. The system begins in the equal superposition over the vertices:
\[ | \psi(0) \rangle = | s_v \rangle, \]
and evolves by Schr\"odinger's equation 
\[ i \frac{\rmd | \psi \rangle}{\rmd t} = H | \psi \rangle \]
with Hamiltonian
\[ H = -\gamma A - | w \rangle \langle w |, \]
where $\gamma$ is the jumping rate (\textit{i.e.}, amplitude per time), $A$ is the adjacency matrix of the graph ($A_{ij} = 1$ if $i$ and $j$ are adjacent and $0$ otherwise), and $| w \rangle$ is the marked vertex we are looking for. The first term effects a quantum walk \cite{CG2004} while the second term acts as an oracle \cite{Mochon2007}, with $\gamma$ setting their relative strength. 

\begin{figure}
\begin{center}
	\includegraphics{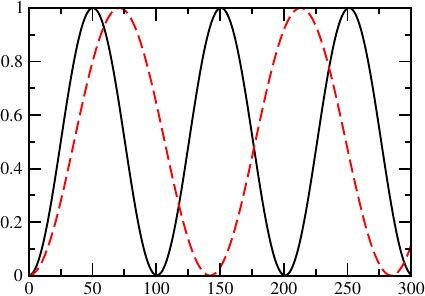}
	\caption{\label{fig:continuous_l0} Success probability as a function of time for continuous-time search on the complete graph with $N = 1024$ (solid black) and $2048$ (dashed red) vertices, at the critical $\gamma = 1/N$.}
\end{center}
\end{figure}

As shown in \fref{fig:complete}, there are only two types of vertices: the marked vertex and the unmarked vertices. So we take equal superpositions of them to be basis vectors of a 2D subspace:
\begin{eqnarray*}
	| a \rangle = | {\rm red} \rangle \\
	| b \rangle = \frac{1}{\sqrt{N-1}} \sum_{i \in {\rm white}} | i \rangle.
\end{eqnarray*}
In this $\{ | a \rangle, | b \rangle \}$ basis, the initial state is
\[ | \psi(0) \rangle = | s_v \rangle = \frac{1}{\sqrt{N}} | a \rangle + \sqrt{\frac{N-1}{N}} | b \rangle, \]
and the Hamiltonian is
\begin{equation}
	\label{eq:H_noloops}
	H = -\gamma \left( \! \begin{array}{cc}
		\frac{1}{\gamma} & \sqrt{N-1} \\
		\sqrt{N-1} & N-2 \\
	\end{array} \right).
\end{equation}
When $\gamma$ takes its critical value of $1/N$ \cite{CG2004,JMW2014,Wong2014}, evolving by Schr\"odinger's equation with this Hamiltonian yields the success probability shown in \fref{fig:continuous_l0}. We see that it reaches a maximum value of $1$, and the runtime (which should be Grover's $\Theta(\sqrt{N})$) scales better than linear (classical) since doubling $N$ less than doubles the runtime.

To show this analytically, note that the eigenvectors of the Hamiltonian with $\gamma = 1/N$ are $\left| \psi_\pm \right\rangle \propto | s_v \rangle \mp | a \rangle$ with corresponding eigenvalues $E_\pm = -1 \pm 1/\sqrt{N} + 1/N$. Then the initial state is $| \psi(0) \rangle = \left( \left| \psi_+ \right\rangle + \left| \psi_- \right\rangle \right)/\sqrt{2}$, and the marked vertex is $| a \rangle = \left( -\left| \psi_+ \right\rangle + \left| \psi_- \right\rangle \right)/\sqrt{2}$. Since the Hamiltonian is time-independent, solving Schr\"odinger's equation yields
\begin{eqnarray*}
	| \psi(t) \rangle
		&= \rme^{-\rmi Ht} | \psi(0) \rangle \\
		&= \rme^{-\rmi Ht} \frac{1}{\sqrt{2}} \left( \left| \psi_+ \right\rangle + \left| \psi_- \right\rangle \right) \\
		&= \frac{1}{\sqrt{2}} \left( \rme^{-\rmi E_+t} \left| \psi_+ \right\rangle + \rme^{-\rmi E_-t} \left| \psi_- \right\rangle \right) \\
		&= \rme^{-\rmi E_-t} \frac{1}{\sqrt{2}} \left( \rme^{-\rmi \Delta Et} \left| \psi_+ \right\rangle + \left| \psi_- \right\rangle \right),
\end{eqnarray*}
where $\Delta E = E_+ - E_-$. When $\Delta E\, t = \pi$, \textit{i.e.},
\[ t = \frac{\pi}{\Delta E} = \frac{\pi}{2} \sqrt{N}, \]
this becomes
\begin{eqnarray*}
	| \psi(t) \rangle 
		&= \rme^{-\rmi E_-\pi/\Delta E} \frac{1}{\sqrt{2}} \left( -\left| \psi_+ \right\rangle + \left| \psi_- \right\rangle \right)
		&= \rme^{-\rmi E_-\pi/\Delta E} | a \rangle.
\end{eqnarray*}
Thus the system evolves to the marked vertex with probability $1$ in time $\pi\sqrt{N}/2$, which agrees with \fref{fig:continuous_l0} with $N = 1024$ and $2048$; as expected, the success probability reaches $1$ at time $\pi\sqrt{1024}/2 \approx 50.265$ and $\pi\sqrt{2048}/2 \approx 71.086$.


\section{Discrete-Time Quantum Walk With Self-Loops}

Now we include $l>0$ self-loops at each vertex, as shown in \fref{fig:complete_loops}. As before, there are only two types of vertices, $a$ and $b$. But now the $a$ vertex can also point towards itself, so the system evolves in a 4D subspace spanned by equal superpositions of the vertices/directions:
\begin{eqnarray*}
	| aa \rangle = | a \rangle \otimes \frac{1}{\sqrt{l}} | a \to a \rangle \\
	| ab \rangle = | a \rangle \otimes \frac{1}{\sqrt{N-1}} \sum_b | a \to b \rangle \\
	| ba \rangle = \frac{1}{\sqrt{N-1}} \sum_b | b \rangle \otimes | b \to a \rangle \\
	| bb \rangle = \frac{1}{\sqrt{N-1}} \sum_b | b \rangle \otimes \frac{1}{\sqrt{N+l-2}} \sum_{b' \sim b} | b \to b' \rangle.
\end{eqnarray*}
In this $\{ | aa \rangle, | ab \rangle, | ba \rangle, | bb \rangle \}$ basis, the initial state is
\begin{eqnarray*}
	| \psi_0 \rangle = \frac{1}{\sqrt{N(N+l-1)}} \Big( 
		&\sqrt{l} | aa \rangle + \sqrt{N-1} | ab \rangle + \sqrt{N-1} | ba \rangle \\
		&+ \sqrt{(N-1)(N+l-2)} | bb \rangle \Big).
\end{eqnarray*}
With self-loops, the $C_1^{\rm flip} = -C_0$ and $C_1^{\rm SKW} = -I_d$ coins now result in different search operators \eref{eq:U} and evolutions, which we analyze separately.

With $C_1^{\rm flip} = -C_0$, \cite{AKR2005} showed that with this coin and $l = 1$ self-loop at each vertex, two applications of the search operator \eref{eq:Uflip} corresponds exactly to Grover's iterate \eref{eq:Grover} on the vertex space. Reproducing their argument, the vertex and coin spaces have equal dimension $N$ in this case. Then $C_0 = R_{s^\perp} = 2 | s \rangle \langle s | - I_N$ is the reflection about the equal superposition $| s \rangle$ over the $N$-dimensional computational basis in Grover's algorithm. With this substitution, the search operator \eref{eq:Uflip} becomes $U = S \cdot (I_N \otimes R_{s^\perp}) \cdot (R_w \otimes I_N)$. Acting by this on the initial equal superposition state $| \psi_0 \rangle = | s \rangle \otimes | s \rangle$, we get $U | \psi_0 \rangle = S (R_w | s \rangle \otimes R_{s^\perp} | s \rangle) = R_{s^\perp} | s \rangle \otimes R_w | s \rangle$. Acting a second time, $U^2 | \psi_0 \rangle = S(R_w R_{s^\perp} | s \rangle \otimes R_{s^\perp} R_w | s \rangle) = R_{s^\perp} R_w | s \rangle \otimes R_w R_{s^\perp} | s \rangle$, which is precisely Grover's iterate \eref{eq:Grover} on the first tensor factor. Thus the success probability reaches $1$ in $\pi \sqrt{N}/2$ applications of $U$, which is an improvement over the success probability of $1/2$ without any self-loops.

\begin{figure}
\begin{center}
	\includegraphics{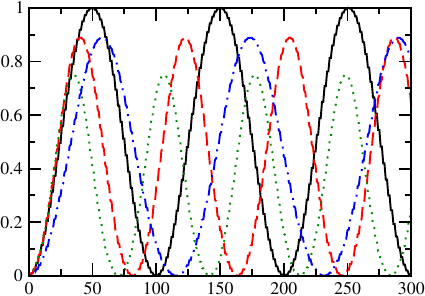}
	\caption{\label{fig:discrete_flip_loops} Success probability as a function of the number of discrete-time applications of $U$ with the $C_1^{\rm flip}$ coin for search on the complete graph with $N$ vertices and $l$ self-loops at each vertex. The solid black, dashed red, and dotted green curves correspond to $N = 1024$ with $l = 1$, $2$, and $3$, respectively, and the dot-dashed blue curve corresponds to $N = 2048$ with $l = 2$.}
\end{center}
\end{figure}

To find the behavior of the algorithm for general $l > 0$, consider the search operator \eref{eq:U} or \eref{eq:Uflip}, which in the $\{ | aa \rangle, | ab \rangle, | ba \rangle, | bb \rangle \}$ basis is
\begin{equation}
	\label{eq:U_loop_flip}
	U = \left( \!\! \begin{array}{cccc}
		\cos\theta & -\sin\theta & 0 & 0 \\
		0 & 0 & -\cos\phi & \sin\phi \\
		-\sin\theta & -\cos\theta & 0 & 0 \\
		0 & 0 & \sin\phi & \cos\phi \\
	\end{array} \!\! \right),
\end{equation}
where $\theta$ is defined such that
\[ \cos\theta = \frac{N-l-1}{N+l-1}, \quad {\rm and} \quad \sin\theta = \frac{2\sqrt{l(N-1)}}{N+l-1}, \]
and $\phi$ is defined such that
\[ \cos\phi = \frac{N+l-3}{N+l-1}, \quad {\rm and} \quad \sin\phi = \frac{2\sqrt{N+l-2}}{N+l-1}. \]
Repeatedly applying this to the initial state, the success probability evolves as shown in \fref{fig:discrete_flip_loops}. Beyond $l = 1$, additional self-loops causes the buildup of success probability to stall, resulting in a lower maximum success probability. The figure also reveals that the maximum success probability only depends on $l$, which is reasonable since the $l = 0$ and $l = 1$ cases achieve success probabilities of $1/2$ and $1$, independent of $N$.

To find the precise behavior of the algorithm, we find the eigenvalues and eigenvectors of the search operator $U$. The eigenvalues are
\[ -1, 1, \rme^{-\rmi\alpha}, \rme^{\rmi\alpha}, \]
where
\[ \cos\alpha = \frac{\cos\theta + \cos\phi}{2} = \frac{N-2}{N+l-1}, \]
\[ \sin\alpha = \frac{\sqrt{(2+\cos\theta+\cos\phi)(2-\cos\theta-\cos\phi)}}{2} = \frac{\sqrt{(2N+l-3)(l+1)}}{N+l-1}, \]
and the corresponding (unnormalized) eigenvectors are
\[ \phi_{-1} = \left( -\frac{\sin\theta}{1+\cos\theta} \frac{1+\cos\phi}{\sin\phi}, -\frac{1+\cos\phi}{\sin\phi}, -\frac{1+\cos\phi}{\sin\phi}, 1 \right)^\intercal \]
\[ \phi_{1} = \left( -\frac{1+\cos\theta}{\sin\theta} \frac{\sin\phi}{1+\cos\phi}, \frac{\sin\phi}{1+\cos\phi}, \frac{\sin\phi}{1+\cos\phi}, 1 \right)^\intercal \]
\[ \phi_{-\alpha} = \left( \frac{\sin\theta}{\sin\phi}, \frac{\cos\theta - \cos\phi}{2\sin\phi} + i \frac{\sin\alpha}{\sin\phi}, \frac{\cos\theta - \cos\phi}{2\sin\phi} - i \frac{\sin\alpha}{\sin\phi}, 1 \right)^\intercal \]
\[ \phi_{+\alpha} = \left( \frac{\sin\theta}{\sin\phi}, \frac{\cos\theta - \cos\phi}{2\sin\phi} - i \frac{\sin\alpha}{\sin\phi}, \frac{\cos\theta - \cos\phi}{2\sin\phi} + i \frac{\sin\alpha}{\sin\phi}, 1 \right)^\intercal. \]
Then
\begin{eqnarray*}
	&2 \frac{1+\cos\phi}{1+\cos\theta} \frac{\sin^2\theta}{\sin^2\phi} \phi_{1} + \phi_{+\alpha} + \phi_{-\alpha} \\
	&\quad = \left( 0, 2\frac{1-\cos\alpha}{\sin\phi}, 2\frac{1-\cos\alpha}{\sin\phi}, 4 \frac{1-\cos\alpha}{1-\cos\phi} \right)^\intercal
\end{eqnarray*}
is dominated by the last term because $\sin\phi \approx \phi$ in the denominator of the second and third terms is small for large $N$, which implies that $1 - \cos\phi \approx \phi^2/2$ in denominator of the last term. Thus if we normalize it to leading-order,
\[ \frac{1-\cos\phi}{4(1-\cos\alpha)} \left( 2 \frac{1+\cos\phi}{1+\cos\theta} \frac{\sin^2\theta}{\sin^2\phi} \phi_{1} + \phi_{+\alpha} + \phi_{-\alpha} \right) \approx \left( 0, 0, 0, 1 \right)^\intercal = | bb \rangle. \]
Note that the initial state $| \psi_0 \rangle \approx | bb \rangle$. Then after $t$ applications of $U$, the system is in the state
\[ U^t \left| \psi_0 \right\rangle \approx \frac{1-\cos\phi}{4(1-\cos\alpha)} \left( 2 \frac{1+\cos\phi}{1+\cos\theta} \frac{\sin^2\theta}{\sin^2\phi} \phi_{1} + \rme^{\rmi \alpha t} \phi_{+\alpha} + \rme^{-\rmi \alpha t} \phi_{-\alpha} \right). \]
We choose $t$ such that $\alpha t = \pi$, \textit{i.e.}, the runtime is
\[ t = \frac{\pi}{\alpha} \approx \left\{ \begin{array}{ll}
	{\pi \over \sqrt{2(l+1)}} \sqrt{N} & l = o(N) \\
	\pi / \sin^{-1} \left( {\sqrt{c(c+2)} \over {c+1}} \right) & l = cN \\
	2 & l = \omega(N) \\
\end{array} \right. \]
for large $N$. At this runtime, the state of the system is
\begin{eqnarray*}
	U^t \left| \psi_0 \right\rangle 
		&\approx \frac{1-\cos\phi}{4(1-\cos\alpha)} \Bigg( &2 \frac{1+\cos\phi}{1+\cos\theta} \frac{\sin^2\theta}{\sin^2\phi} \phi_{1} - \phi_{+\alpha} - \phi_{-\alpha} \Bigg) \\
		&= \frac{1-\cos\phi}{4(1-\cos\alpha)} \Bigg( 
			&\!-4 \frac{\sin\theta}{\sin\phi}, \frac{2-3\cos\theta+\cos\phi}{\sin\phi}, \\
			&&\frac{2-3\cos\theta+\cos\phi}{\sin\phi}, 2\frac{-\cos\theta + \cos\phi}{1-\cos\phi} \Bigg)^\intercal.
\end{eqnarray*}
Then the success probability $p$ is given by the sum of the squares of the first two terms:
\[ p = \frac{(1-\cos\phi)^2}{16(1-\cos\alpha)^2} \left( \frac{16\sin^2\theta + (2-3\cos\theta+\cos\phi)^2}{\sin^2\phi} \right). \]
Plugging in for $\cos\alpha$, $\sin\phi$, $\cos\phi$, $\sin\theta$, and $\cos\theta$, this is
\[ p = \frac{16l(N-1) + (3l-1)^2}{4 (l+1)^2 (N+l-2)} \approx \left\{ \begin{array}{ll}
	{4l \over (l+1)^2} & l = o(N) \\
	{16+9c \over 4c(c+1)} {1 \over N} & l = cN \\
	{9 \over 4l} & l = \omega(N) \\
\end{array} \right. \]
for large $N$. These expressions for $t$ and $p$ agree with \fref{fig:discrete_flip_loops}; for search with $N = 1024$ vertices and $l = 1,2,3$, the runtimes are respectively $\pi\sqrt{1024}/\sqrt{2(l+1)} \approx 50, 41, 36$ with corresponding success probabilities $4l/(l+1)^2 = 1, 0.889, 0.75$. With $N = 2048$ and $l = 2$, the runtime is $\pi\sqrt{2048}/\sqrt{2(2+1)} \approx 58$ with success probability $4(2)/(2+1)^2 \approx 0.889$.

These results indicate that the maximum success probability decreases as $l$ increases. Even so, there is still an improvement over the loopless success probability of $1/2$, so long as $l \le 5$. For $l$ beyond this, the success probability is less than $1/2$, but if $l$ is any constant, the overall runtime is still Grover's $\Theta(\sqrt{N})$. When $l$ scales greater than a constant, the quadratic quantum speedup is lost due to the classical repetitions of the algorithm needed to boost the success probability. Despite this, when $l = o(N)$, we still obtain an improvement over the classical algorithm's $\Theta(N)$ runtime. Finally, when $l = \Omega(N)$, the success probability fails to increase beyond its initial scaling of $\Theta(1/N)$, and so the quantum algorithm is no better than the classical one.

\begin{figure}
\begin{center}
	\includegraphics{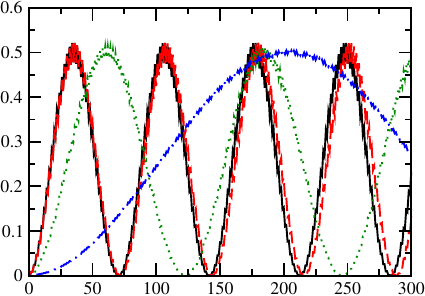}
	\caption{\label{fig:discrete_AKR_N1024_loops} Success probability as a function of the number of discrete-time applications of $U$ with the $C_1^{\rm SKW}$ coin for search on the complete graph with $N = 1024$ vertices and $l$ self-loops at each vertex. The solid black, dashed red, dotted green, and dot-dashed blue curves correspond to $l = 0$, $\sqrt{N} = 32$, $2N = 2048$, and $N^{3/2} = 32768$, respectively.}
\end{center}
\end{figure}

Now consider $C_1^{\rm SKW} = -I_d$. The search operator \eref{eq:U} in the $\{ | aa \rangle, | ab \rangle, | ba \rangle, | bb \rangle \}$ basis is
\begin{equation}
	\label{eq:U_loop_SKW}
	U = \left( \!\! \begin{array}{cccc}
		-1 & 0 & 0 & 0 \\
		0 & 0 & -\cos\theta & \sin\theta \\
		0 & -1 & 0 & 0 \\
		0 & 0 & \sin\theta & \cos\theta \\
	\end{array} \!\! \right),
\end{equation}
where
\[ \cos\theta = \frac{N+l-3}{N+l-1}, \quad {\rm and} \quad \sin\theta = \frac{2\sqrt{N+l-2}}{N+l-1}. \]
Clearly, $| aa \rangle$ is an eigenvector of $U$ with eigenvalue $-1$. The remaining part of $U$ corresponding to $| ab \rangle$, $| ba \rangle$, and $| bb \rangle$ takes the same form as $U$ for the $l = 0$ case \eref{eq:U_noloops}. Since $\theta$ is small for large $N$, those results carry over: the success probability reaches $1/2$ in
\[ t = \frac{\pi}{2 \sin^{-1} \left( \frac{\sqrt{(1-\cos\theta)(3+\cos\theta)}}{2} \right)} \]
applications of $U$. The scaling of this with $N$ depends on the scaling of $l$. In particular, for large $N$,
\[ t = \left\{ \begin{array}{ll}
	{\pi \over 2\sqrt{2}} \sqrt{N} & l = o(N) \\
	\case{\pi \sqrt{c+1}}{2\sqrt{2}} \sqrt{N} & l = cN \\
	\case{\pi}{2\sqrt{2}} \sqrt{l} & l = \omega(N) \\
\end{array} \right. . \]
An example of this evolution is shown in \fref{fig:discrete_AKR_N1024_loops}, with the success probability reaching the expected $1/2$ at time $\pi\sqrt{1024}/2\sqrt{2} \approx 36$ for both $l = 0$ and $l = \sqrt{N} = 32$, time $\pi\sqrt{1+2}\sqrt{1024}/2\sqrt{2} \approx 62$ for $l = 2N = 2048$, and time $\pi\sqrt{32768}/2\sqrt{2} \approx 201$ for $l = N^{3/2} = 32768$.

With this coin, self-loops cause the success probability to take longer to reach $1/2$. As long as the number of self loops scales less than or equal to $N$ (\textit{i.e.}, $l = O(N)$), the runtime still scales as Grover's $\Theta(\sqrt{N})$. Furthermore, there is still a speedup over the classical algorithm so long as $l$ scales less than $N^2$ (\textit{i.e.}, $l = o(N^2)$). Compared to the $C_1^{\rm flip}$ coin, which obtains Grover's $\Theta(\sqrt{N})$ runtime when $l = \Theta(1)$ and a speedup over classical so long as $l = o(N)$, this $C_1^{\rm SKW}$ coin is in some sense more robust to self-loops; it takes more of them for the walk to lose its quantum speedup.


\section{Continuous-Time Quantum Walk With Self-Loops}

Let us see how the continuous-time quantum walk algorithm is affected by the presence of $l$ self-loops at each vertex. If we count each self-loop to contribute $1$ to the diagonal of the full $N$-dimensional adjacency matrix so that $A_{ii} = l$ at vertex $i$ \footnote{Note that some treatments count each self-loop as 2 in the adjacency matrix, which would result in $A_{ii} = 2l$, but it makes no difference to our result.}, then in the two-dimensional subspace spanned by $\{ | a \rangle, | b \rangle \}$, the Hamiltonian is
\[ H = -\gamma \left( \! \begin{array}{cc}
	\frac{1}{\gamma} + l& \sqrt{N-1} \\
	\sqrt{N-1} & N+l-2 \\
\end{array} \right). \]
Note this is simply the Hamiltonian with no self-loops \eref{eq:H_noloops} plus $l I$. Adding a multiple of the identity matrix to the Hamiltonian in this manner constitutes a rezeroing of energy or an overall phase, so it has no observable effects. Thus the self-loops do not change the evolution at all; with or without self-loops, at the critical $\gamma = 1/N$, the success probability reaches $1$ at time $\pi\sqrt{N}/2$. Thus the continuous-time quantum walk algorithm is completely robust to lackadaisical errors in our model using $l$ self-loops at each vertex.


\section{Generalization to Multiple Marked Vertices}

All of these results are straightforward to generalize to the case of $k$ marked vertices. We assume that $k = o(N)$ since the number of marked vertices cannot scale more than the number of vertices, and if $k = cN$, then one can classically find a marked vertex in a constant number of guesses. The classical search would take an expected $\Theta(N/k)$ time to find one of the $k$ marked vertices on the complete graph of $N$ vertices. As for the quantum walk, let us consider each of the cases above.

\begin{figure}
\begin{center}
	\includegraphics{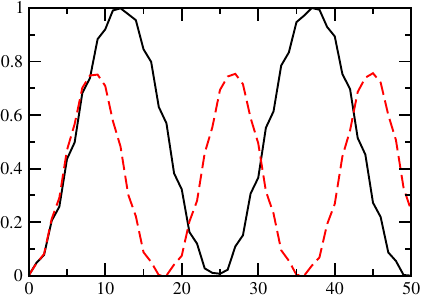}
	\caption{\label{fig:discrete_flip_N1024_k16} For search on the complete graph with $N = 1024$ vertices, $k = 16$ marked vertices, and $l$ self-loops at each vertex, the success probability as a function of the number of discrete-time applications of $U$ with the $C_1^{\rm flip}$ coin. The solid black and dashed red curves are respectively $l = 1$ and $32$.}
\end{center}
\end{figure}

Beginning with discrete-time quantum walks, with $k > 1$ marked vertices and $l$ self-loops at each vertex, the system evolves in a 4D subspace spanned by $\{ | aa \rangle, | ab \rangle, | ba \rangle, | bb \rangle \}$. With the $C_1^{\rm flip} = -C_0$ coin, the search operator \eref{eq:U} or \eref{eq:Uflip} is
\[ U = \left( \!\! \begin{array}{cccc}
	\cos\theta & -\sin\theta & 0 & 0 \\
	0 & 0 & -\cos\phi & \sin\phi \\
	-\sin\theta & -\cos\theta & 0 & 0 \\
	0 & 0 & \sin\phi & \cos\phi \\
\end{array} \!\! \right), \]
where $\theta$ is defined such that
\[ \cos\theta = \frac{N-2k-l+1}{N+l-1}, \quad {\rm and} \quad \sin\theta = \frac{2\sqrt{(N-k)(k+l-1)}}{N+l-1} \]
and $\phi$ is defined such that
\[ \cos\phi = \frac{N-2k+l-1}{N+l-1}, \quad {\rm and} \quad \sin\phi = \frac{2\sqrt{k(N-k+l-1)}}{N+l-1}. \]
This has the same form as the case of one marked vertex \eref{eq:U_loop_flip}, and since $\phi$ is small for large $N$, the solutions carry over: define $\alpha$ such that
\[ \cos\alpha = \frac{\cos\theta + \cos\phi}{2} = \frac{N-2k}{N+l-1} \]
\begin{eqnarray*}
	\sin\alpha 
		&= \frac{\sqrt{(2+\cos\theta+\cos\phi)(2-\cos\theta-\cos\phi)}}{2} \\
		&= \frac{\sqrt{(2N-2k+l-1)(2k+l-1)}}{N+l-1}.
\end{eqnarray*}
Then after
\[ t = \frac{\pi}{\alpha} \approx \left\{ \begin{array}{ll}
	{\pi \over \sqrt{2(2k+l-1)}} \sqrt{N} & l = o(N) \\
	\pi / \sin^{-1} \left( {\sqrt{c(c+2)} \over {c+1}} \right) & l = cN \\
	2 & l = \omega(N) \\
\end{array} \right. \]
applications of $U$, the success probability reaches a maximum value of
\begin{eqnarray*}
	p 
		&= \frac{(1-\cos\phi)^2}{16(1-\cos\alpha)^2} \left( \frac{16\sin^2\theta + (2-3\cos\theta+\cos\phi)^2}{\sin^2\phi} \right) \\
		&= \frac{k \left[ 16N(k+l-1) + 9(l-1)^2 - 4k(l-1) - 12k^2 \right]}{4(2k+l-1)^2 (N-k+l-1)}  \\
		&\approx \left\{ \begin{array}{ll}
			{4k(k+l-1) \over (2k+l-1)^2} & l = o(N) \\
			{16+9c \over 4c(c+1)} {k \over N} & l = cN \\
			{9k \over 4l} & l = \omega(N) \\
		\end{array} \right. .
\end{eqnarray*}
This is shown in \fref{fig:discrete_flip_N1024_k16}, where the success probability reaches $1$ at $\pi\sqrt{1024}/\sqrt{2(2\cdot16+1-1} \approx 13$ and $4\cdot16(16+32-1)/(2\cdot16+32-1)^2 \approx 0.758$ at $\pi\sqrt{1024}/\sqrt{2(2\cdot16+32-1} \approx 9$ applications of $U$, as expected.

\begin{figure}
\begin{center}
	\includegraphics{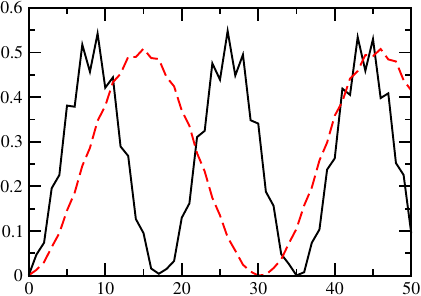}
	\caption{\label{fig:discrete_AKR_N1024_k16} Success probability as a function of the number of discrete-time applications of $U$ with the $C_1^{\rm flip}$ coin for search on the complete graph with $N = 1024$ vertices, $k = 16$ marked vertices, and $l$ self-loops at each vertex. The solid black and dashed red curves are respectively $l = 1$ and $32$.}
\end{center}
\end{figure}

With the $C_1^{\rm SKW} = -I_d$ coin, the search operator \eref{eq:U} in this basis is
\[ U = \left( \!\! \begin{array}{cccc}
	-1 & 0 & 0 & 0 \\
	0 & 0 & -\cos\theta & \sin\theta \\
	0 & -1 & 0 & 0 \\
	0 & 0 & \sin\theta & \cos\theta \\
\end{array} \!\! \right), \]
where
\[ \cos\theta = \frac{N-2k+l-1}{N+l-1}, \quad {\rm and} \quad \sin\theta = \frac{2\sqrt{k(N-k+l-1)}}{N+l-1}. \]
This has the same form as the case of one marked vertex \eref{eq:U_loop_SKW}, and since $\theta$ is small for large $N$, the solutions carry over: we reach a success probability of $1/2$ in
\[ t = \frac{\pi}{2 \sin^{-1} \left( \frac{\sqrt{(1-\cos\theta)(3+\cos\theta)}}{2} \right)} = \left\{ \begin{array}{ll}
	{\pi \over 2\sqrt{2k}} \sqrt{N} & l = o(N) \\
	\case{\pi \sqrt{c+1}}{2\sqrt{2k}} \sqrt{N} & l = cN \\
	\case{\pi}{2\sqrt{2k}} \sqrt{l} & l = \omega(N) \\
\end{array} \right. \]
applications of $U$. This is shown in \fref{fig:discrete_AKR_N1024_k16}, where the success probability reaches $1/2$ at $\pi\sqrt{1024}/2\sqrt{2\cdot16} \approx 9$ and $\pi\sqrt{2+1}\sqrt{1024}/2\sqrt{2\cdot16} \approx 15$ applications of $U$, as expected.

Finally for the continuous-time quantum walk, the Hamiltonian is
\[ H = -\gamma \left( \begin{array}{cc}
	\frac{1}{\gamma} + k+l-1 & \sqrt{k(N-k)} \\
	\sqrt{k(N-k)} & N-k+l-1 \\
\end{array} \right). \]
This simply adds $lI$ to the Hamiltonian with no self-loops \cite{Wong2015}, which is a rezeroing of energy or global phase, so it has no observable effects.


\section{Conclusion}

We have proposed a quantum analogue of classical lazy random walks, called lackadaisical quantum walks, where the quantum walker has some preference to stay put by introducing $l$ self-loops at each vertex of the graph. We have investigated the consequences of this for quantum search on the complete graph, showing the self-loops can have vastly different effects on quantum search depending on the type of quantum walk. For discrete-time quantum walks with the $C_1^{\rm flip}$ coin, one self-loop per vertex boosts the success probability, but additional self-loops hinders it. With the $C_1^{\rm SKW}$ coin, however, rather than hindering the maximum success probability, the self-loops slow down its buildup. This hindrance is less potent, eliminating all speedup over classical search when $l = \omega(N^2)$ compared to the first coin's $l = \omega(N)$. Continuous-time quantum walks, on the other hand, are not affected at all by the self-loops. All of these results extend to multiple marked vertices.


\ack
This work was supported by the European Union Seventh Framework Programme (FP7/2007-2013) under the QALGO (Grant Agreement No.~600700) project, and the ERC Advanced Grant MQC.


\section*{References}
\bibliographystyle{iopart-num}
\bibliography{refs}


\includepdf[pages={1-2}]{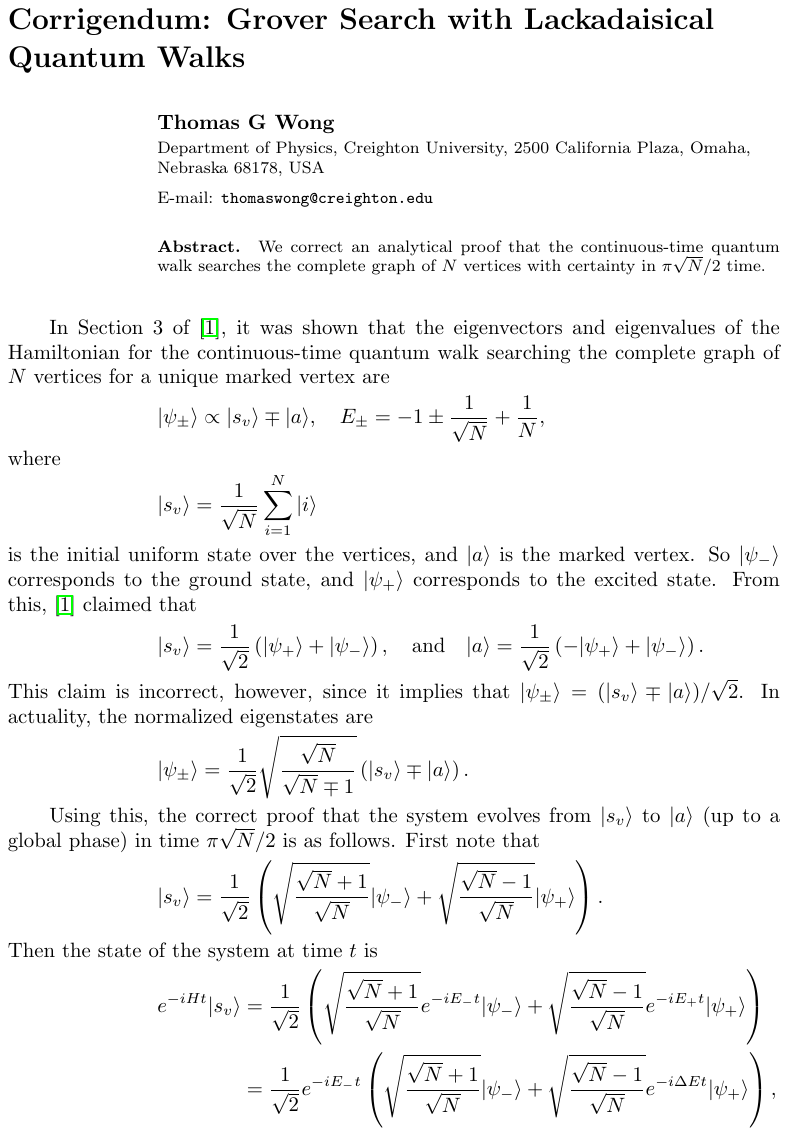}

\end{document}